\documentstyle[aps,pra,epsfig]{revtex}

\newcommand{\be}{\begin{equation}} 
\newcommand{\en}{\end{equation}} 
\newcommand{\bea}{\begin{eqnarray}} 
\newcommand{\ena}{\end{eqnarray} } 
\newcommand{\grs}{\biggl\{} 
\newcommand{\grd}{\biggr\}} 
\newcommand{\ii}{{\bf i}} 
\newcommand{\jj}{{\bf j}} 
\newcommand{\kk}{{\bf k}} 
\newcommand{\rr}{{\bf r}} 
\newcommand{\es}{{\bf s}} 
\newcommand{\oo}{{\bf 0}} 
\newcommand{\pp}{{\bf p}} 
\newcommand{\intphi}{\int \prod_{\ii} D 
 \varphi_{\ii} \: \:} 
\newcommand{\intau}{\int_{0}^{\beta} d\tau } 
\newcommand{\intpsi}{\int \prod_{\ii}D^2  \psi_{\ii} \: \:} 
\newcommand{\sumij}{\sum_{\ii \jj}} 
\newcommand{\sumi}{\sum_{\ii}}

\newcommand{\psii}{\psi_{\ii}} 
\newcommand{\psij}{\psi_{\jj}} 
\newcommand{\varphii}{\varphi_{\ii}} 
\newcommand{\varphij}{\varphi_{\jj}} 
\newcommand{\varphir}{\varphi_{\rr}} 
\newcommand{\varphis}{\varphi_{\es}} 
\newcommand{\psir}{\psi_{\rr}} 
\newcommand{\psis}{\psi_{\es}} 
\newcommand{\phii}{\phi_{\ii}} 
\newcommand{\phij}{\phi_{\jj}}

\begin{document}
\title{On the Phase Diagram of Josephson Junction Arrays
with Offset Charges\footnote{in
{\em
New Developments in Superconductivity Research}, R. S.
Stevens ed., pp. 31-52, Nova Science Publishers, New York (2003) ISBN
1-59033-862-6.}}
\author{F.P. Mancini$^1$, P. Sodano$^1$, and A. Trombettoni$^{2}$}
\address{$^1$Dipartimento di Fisica and Sezione I.N.F.N., Universit\`a di
Perugia,\\ Via A. Pascoli, I-06123 Perugia, Italy}
\address{$^{2}$ I.N.F.M. and Dipartimento di Fisica, Universit\`a
di Parma, \\ parco Area delle Scienze 7A, I-43100, Parma, Italy}
\date{\today} 
\maketitle 
 
\begin{abstract} 
We study the effects of external offset charges on  
the phase diagram of Josephson junction arrays.
Using the path integral approach, we provide a pedagogical
derivation of the equation for the phase boundary line 
between the insulating and the superconducting phase within the
mean-field theory approximation. For a uniform offset
charge $q=e$ the superconducting phase increases with respect to $q=0$
and a characteristic lobe structure appears in the phase diagram when
the critical line is plotted as a function of $q$ at fixed temperature.
We review our analysis of the physically relevant situation where
a Josephson network feels the effect of random offset charges. We
observe that the Mott-insulating lobe structure of the phase diagram
disappears for large variance ($\sigma \gtrsim e$) of the offset charges
probability distribution; with nearest-neighbor interactions, the
insulating lobe around $q=e$ is destroyed even for small values of
$\sigma$. Finally, we study the case of random self-capacitances: here
we observe that, until the variance of the distribution reaches a
critical value, the superconducting phase increases in comparison to the
situation in which all self-capacitances are equal.
\end{abstract}
\pacs{PACS: 74.25.Dw, 05.30.Jp, 74.50.+r, 85.25.Cp}

\section{INTRODUCTION}
Josephson junction arrays (JJA) are an ideal model to study
a variety of phenomena such as phase transitions, 
frustration effects, and vortex dynamics \cite{fazio01}.
The high level of accuracy reached 
in their experimental realization makes JJA relevant
for different 
applications, ranging from possible implementations 
of quantum computation schemes \cite{makhlin01} 
to the realization of topological order \cite{doucot03}. 
Furthermore, 
particular geometries of JJA may induce novel
classical  coherent states \cite{burioni00}. 

JJA and granular superconductors, namely systems of metallic grains
embedded in an insulator, become superconducting in two steps
\cite{simanek94}. First, at the bulk critical temperature,
each grain develops a superconducting gap but the phases 
of the order parameter on different grains are uncorrelated. 
Then, at a lower temperature $T_c$, the Cooper pair tunneling between
grains gives rise to a long-range phase coherence and the system as a
whole exhibits a phase transition to a superconducting state. The
latter transition is governed by the competition between the Josephson
tunneling, characterized by a Josephson coupling energy $E_J$, and the
Coulomb interaction between Cooper pairs, described by a charging energy
$E_C$. In classical junction arrays the Josephson coupling $E_J$ is
dominant and the transition separates a superconducting low temperature
phase from a normal high temperature phase. When $E_C$ is comparable to
$E_J$ (small grains), charging effects give rise to a quantum dynamics
and the energy cost of Cooper pair tunneling may be higher then the
energy gained by the formation of a phase-coherent state.

It is relevant to analyze in detail the effect of a background of
external charges on the superconductor-insulator transition of a quantum
JJA. Offset  charges arise in real physical systems as a result of
charged impurities or by  application of a gate voltage between the
array and the ground.
In the former case, offset charges are distributed randomly on
the lattice while in the latter case they play the role of a sort of
chemical potential and their distribution can be uniform. Thus, offset
charges may be regarded as effective charges $q_{\ii}$, located at the
sites of the lattice. When $q_{\ii}\ne 2e$, the offset charges  cannot
be eliminated by Cooper pair tunneling. As it will be clear, offset
charges frustrate the attempts of the system to minimize the energy of
the charge distribution of the ground state (for this reason they are
also called {\em frustration charges}). A large number of studies has by
now been devoted to the analysis of the effects induced by offset
charges on the (zero-temperature) quantum phase transition
\cite{roddick93,luciano9596,larkin97,choi01} and on the phase transition
at finite $T$ \cite{bruder92,vanotterlo93,grignani00}.
 
The Hamiltonian commonly used to describe the Cooper pair tunneling in 
superconducting quantum networks defines the so called quantum phase model 
(QPM). In its most general form it is given by  
\begin{equation} 
H=\frac{1}{2}\sum_{ij}(Q_{{\bf i}}+q_{{\bf i}})C_{{\bf i}{\bf j}}^{-1}(Q_{
{\bf j}}+q_{{\bf j}}) -E_{J}\sum_{\left \langle ij\right\rangle }\cos 
(\varphi_{{\bf i}}-\varphi_{{\bf j}})  
\label{QPM} 
\end{equation} 
where $\varphi_{{\bf i}}$ is the phase 
of the superconducting order 
parameter at the grain ${\bf i}$. 
Its conjugate variable ${n}_{{\bf i}}$ ($[{
\varphi_{{\bf i}}}, {n}_{{\bf i}}]=
i\:\delta_{{\bf i} {\bf j}}$) describes 
the number of Cooper pairs in the ${\bf i}$-th 
superconducting grain. The 
symbol $\left\langle ij\right\rangle$ indicates a sum over
nearest-neighbor grains only. The first term in the Hamiltonian
(\ref{QPM}) determines the electrostatic coupling between the Cooper
pairs: ${Q}_{{\bf i}}$ is the excess of charge due to Cooper pairs ($
Q_{{\bf i}}=2e n_{{\bf i}}$) on the site ${\bf i}$ and $C_{{\bf i} {\bf
j}}$ is the capacitance matrix. The diagonal elements of the inverse
matrix $C_{\ii\jj}^{-1}$ provide the charging energy: $E_C=e^2
C_{\ii\ii}^{-1}/2\equiv e^2 /2 C_0$, where $C_0$ is the
self-capacitance. The second term describes the hopping of Cooper pairs
between neighboring sites ($E_J$ is the Josephson energy).
The parameter $\alpha=z E_J/4 E_C$ (where $z$ is
the coordination number) governs the superconductor-insulator
transition: for $\alpha \ll 1$ the Josephson network is in the
Mott-insulator state, while for $\alpha \gg 1$ is in the superconducting
one. An external gate voltage $V_{{\bf i}}$ contributes to the energy
via the offset charge $q_{{\bf i}}=\sum_{{\bf j}} C_{{\bf i}{\bf j}}
V_{{\bf j}}$. This external voltage can be either applied
to the ground plane or, more realistically, it may be induced by charges
trapped in the substrate. 
In the latter situation $q_{{\bf i}}$ is naturally a random
variable: we shall review our study of this physically interesting
situation in Section IV.

As it is well known \cite{fazio01}, the QPM (\ref{QPM}) is equivalent to
the boson Hubbard model (BHM) in the limit of large particle numbers per
junction. The BHM describes soft-core bosons hopping on a lattice
\cite {fisher89} and is defined as 
\begin{equation} 
H=\frac{1}{2}\sum_{ij} n_{{\bf i}} 
U_{ij} n_{{\bf j}}- \mu \sum_{i} n_{{\bf i 
}} - t \sum_{\left \langle ij\right\rangle }
(b_{{\bf i}}^{\dag}b_{{\bf j}}+H.C.).  
\label{BHM} 
\end{equation} 
Here, $b_{{\bf i}}^{\dag}$ ($b_{{\bf i}}$) 
is the creation (annihilation) 
operator for bosons and $n_{{\bf i}}=b_{{\bf i}}^{\dag}
b_{{\bf i}}$ is the 
number operator. By writing the $b_{{\bf i}}$'s
in terms of their
amplitudes and phases, and by neglecting the deviations
of the amplitudes from the average number of Cooper pairs  $\langle
n\rangle$, we are lead to the QPM (\ref{QPM}). An exact mapping between
the two models has been derived in Ref. \cite{anglin01}. 
The hopping term is associated with the Josephson tunneling
($2\langle n\rangle t \to E_J$)
whereas $U_{{\bf i}{\bf j}} \to 4e^2 C_{{\bf i}{\bf j}}^{-1}$ 
describes the Coulomb interactions between bosons.
The chemical potential in the BHM plays 
a role analogous to the external charge in the QPM
($\mu \to q_{{\bf i}}$). 
Thus, a QPM with random offset charges corresponds to a BHM with
random on-site energies. 

\begin{figure}[h]
\centerline{\psfig{figure=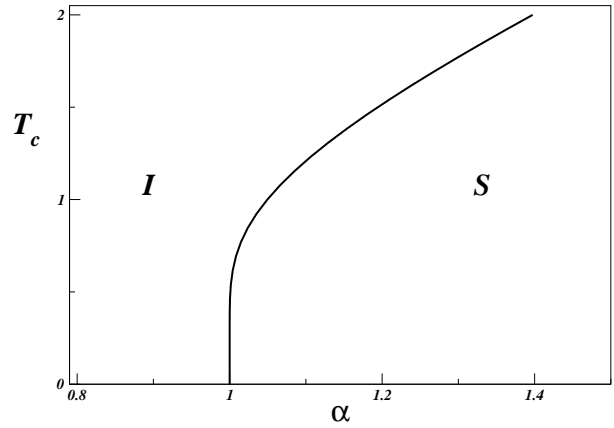,width=72mm,angle=270}}
\caption{Phase diagram for the diagonal model without charge 
frustration. The critical temperature $T_c$ is in units of $k_B /E_C$
and $\alpha$  stands for the ratio $z E_J /4E_C$. The {\bf  I} and
{\bf  S} indicate, respectively, insulating and superconducting
phase.} \label{fig1}
\end{figure}

\begin{figure}[h]
\centerline{\psfig{figure=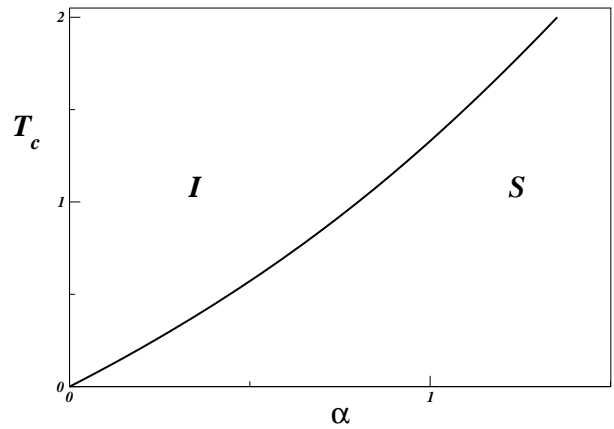,width=72mm,angle=270}}
\caption{Phase diagram of the diagonal model with half-integer 
charge frustration $q=e$; $T_c$ is in units of $k_B /E_C$.}
\label{fig2}
\end{figure}

\begin{figure}[h]
\centerline{\psfig{figure=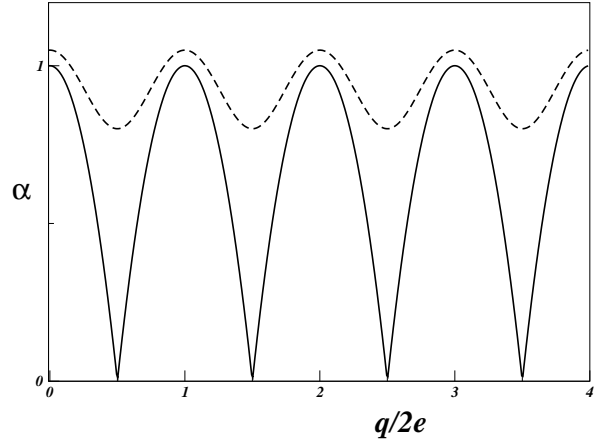,width=72mm,angle=270}}
\caption{Lobe diagram for $T_c \to 0$ (solid line) and $T_c=k_B /E_C$ (dashed
line). As the Hamiltonian (\ref{QPM}) is periodic in $q$ with period
$2e$, the lobes are repeated on the $q$ axis.}
 \label{fig3}
\end{figure}

In this paper we shall use a pedagogical approach to review the results
obtained in our analysis \cite{grignani00,mancini03} of the finite
temperature superconductor-insulator transition in JJA. Using the finite
temperature path integral approach, we provide an explicit derivation of
the equation for the phase boundary for quantum JJA with offset charges
and general capacitance matrix within the mean-field theory
approximation \cite{grignani00,kopec00}.
Offset charges dramatically change the phase boundary: as it may
be easily seen from Figs. \ref{fig1} and \ref{fig2}, already with
diagonal capacitance matrices, and for a uniform offset charge $q=e$,
the superconducting phase increases with respect to the unfrustrated
case. A lobe structure appears in the phase diagram when, at fixed
temperature, the critical line is plotted against the offset charge $q$.
In Fig. \ref{fig3} we plot the value of $\alpha$ at which the transition
occurs as a function of $q$ for two different critical temperatures
$T_c$. As it is shown in Fig. \ref{fig3}, the critical line is
$2e$-periodic, and has a minimum in $q=e$, corresponding to the fact
that for this value of $q$ the superconducting region is maximal.

Within the same mean-field approach, we shall also treat JJA with
capacitive disorder \cite{mancini03}, i.e., random offset charges
and/or random self-capacitances.
This is motivated by the fact that, in practical realizations of
Josephson devices, one has to deal with capacitance disorder caused
either by offset charge defects in the
junctions or in the substrate (random offset charges)
\cite{krupenin00} or
by imperfections in the construction of the devices,
which may lead to
random capacitances of the Josephson junctions.
Although from a theoretical point of view charge and magnetic
frustration are expected
to be dual to each other,
experimentally it is possible to tune only the magnetic
frustration in a controlled way. For this reason,
it is widely believed that a challenging task for the theory is to develop
reliable techniques to investigate the effects of random offset charges
and capacitances on the phase structure of JJA.

Random offset charges could be forced to vanish by using a gate for each
superconducting island; however, this procedure works only for small
networks, since in large arrays too many gate would be necessary, making
impossible the cooling of the system at the desired temperatures. In
Ref.\cite{lafarge95} the case of uniform charge frustration was analyzed
experimentally. It was observed a sensible variation ($\sim 40\%$) of
the resistance between the unfrustrated and the fully frustrated array:
however, it was impossible to quench the Coulomb blockade as it can be
done in small arrays.

During the last decade, much attention have been devoted
to the Bose-Hubbard model with random on-site energies at $T=0$
\cite{fisher89,pazmandi95,kisker97,pazmandi97,herbut98,granato98,lee01}.
As it has been already pointed out, the properties of the disordered BHM
are closely related to those of JJA with random offset charges. Without
disorder, the BHM exhibits two types of phases: a superfluid phase and a
Mott insulating phase. Strong disorder may lead to the existence of a
third, intermediate gapless phase exhibiting an infinite superfluid
susceptibility and finite compressibility: the Bose glass (BG).
It is by now known that, unless one introduces an {\em ad hoc}
probability distribution, there is no BG in mean-field (MF) at $T=0$
\cite{fisher89}; the reason why - at $T=0$ - there is no BG phase in MF
is better understood if one considers the infinite-range hopping limit
of the Hamiltonian (\ref{BHM}): $t_{ij}=t/N$ for all the sites ${\bf i}
\neq {\bf j}$, $N$ being the total number of sites. At $T=0$, when $t=0$
and $\left\langle n_{{\bf i}} \right\rangle=0$ the system is, of course,
in the insulating phase. If a small positive $t$ is turned on, the
infinite-ranged hopping allows the system to gain kinetic energy (with
zero cost in the on-site potential energy) by moving bosons in
unoccupied sites with the highest on-site energies; thus, the bosons
are delocalized and the system becomes superfluid even for very small
values of $t$. Of course, this argument does not hold at finite $T$; so
far, there is no evidence for a Bose glass phase at finite temperature.

In the following we shall review the results of Ref. \cite{mancini03},
where the finite temperature phase diagrams of JJA with random offset
charges and/or random self-capacitances have been obtained.

The plan of the paper is the following: in Section II 
we use the coarse grained approach to compute  
the Ginzburg-Landau free energy for quantum JJA with charge 
frustration and a general Coulomb interaction matrix.
The path integral
providing the phase correlator 
needed to investigate the finite temperature critical behavior of the
system is explicitly computed.

In Section III, within the mean field theory approximation, we derive
the analytical form of the critical line equation from the
Ginzburg-Landau free energy.
The phase diagram  is drawn in the diagonal case for a 
generic uniform offset charge distribution.
We then analyze the low temperature limit of a system 
with nearest neighbor interaction matrix 
when a uniform background of external charges
$q_{\ii}=e$ is considered.  

In Section IV we consider JJA with capacitive disorder
(random offset charges and/or random self-capacitances):
using the results of the previous Sections, we compute the free
energy averaged over the disorder. Section IV.A is devoted to the study of the
effects of random offset charges on the finite temperature phase diagram,
with diagonal and
nearest-neighbor capacitance matrices, while in Section IV.B we consider
JJA with random self-capacitance.

Finally, Section V is
devoted to concluding remarks.

\section{Path Integral Approach: 
the Ginzburg-Landau Free Energy} 

The partition function for the frustrated model
described by the Hamiltonian (\ref{QPM}) is given by 
\be 
\label{msz} 
Z={\rm Tr} e^{-\beta H}=\sum_n \left<\psi_n\left|e^{-\beta H} 
\right|\psi_n\right> 
\en 
where $\beta=1/k_B T$ and the sum is
extended only to states of charge $2e$ and thus with definite 
periodicity. 
 
In the functional approach $Z$ reads 
\be 
Z=\int \prod_{\ii} D\varphi_{\ii} \exp \bigg\{ - 
\int_0^{\beta}d\tau L_E\left(\varphi_{\ii}(\tau), 
\frac{d\varphi_{\ii}}{d\tau}(\tau)\right) \bigg\} 
\en 
where the Euclidean Lagrangian $L_E$  may be derived from
\be \label{11} 
L=\frac{1}{2}\bigg(\frac{\hbar}{2e} \bigg)^2\sum_{\ii \jj}C_{\ii \jj} 
\frac{d\varphi_{\ii}}{dt}   \frac{d\varphi_{\jj}}{dt}-\bigg(\frac{\hbar}{2e} 
\bigg)\sum_{\ii}\frac{d\varphi_{\ii}}{dt}q_{\ii} +E_J
\sum_{\left<\ii\jj\right>} \cos (\varphi_{\ii}-\varphi_{\jj})
\en 
by replacing $i t / \hbar \to \tau$. 
The path integral that one should compute is then given by: 
\be  \label{12} 
Z= \intphi \exp  \bigg\{ \int_{0}^{\beta} d\tau 
\big[ - \frac{1}{2} \sum_{\ii \jj} C_{\ii \jj} 
\frac{\dot{\varphi_{\ii}}}{2e} \frac{\dot{\varphi_{\jj}}}{2e} 
+ i \sum_{\ii}q_{\ii} \frac{\dot{\varphi_{\ii}}}{2e}+ 
\frac{E_J}{2}\sum_{ \ii \: \jj }e^{ i\varphi_{\ii}} 
\gamma_{\ii \jj} e^{-i\varphi_{\jj}} \big]\bigg\} 
\en 
where $-\infty < \varphi_{\ii} < + \infty $, 
$\varphi_{\ii}(0)=\varphi_{\ii}(\beta)+2\pi n_{\ii}$ 
and $\gamma_{\ii \jj}=1$ if $\ii,\jj$ are nearest 
neighbors and equals zero otherwise (i.e., hopping term just between nearest 
neighbors). If $\gamma_{{\bf i} {\bf j}}=1$ for all pairs ${\bf i},{\bf j}$ 
on the lattice, one is lead to the infinite-range hopping limit which 
provides a remarkable example of an exactly solvable MF theory
\cite{fisher89}. The integers $n_{\ii}$ appearing in the
boundary conditions take into account the $2\pi$-periodicity of the 
states $\psi_n$ contributing to Eq. (\ref{msz}).
 
In order to derive the Ginzburg-Landau free energy for the order parameter, 
it is convenient to carry out the integration over the phase 
variables by means of the Hubbard-Stratonovich 
procedure \cite{hubbard59}: using the identity 
\be \label{gauss} 
 e^{J^{+} \Gamma J} = 
\frac{\det \Gamma^{-1}}{\pi^{N}} \intpsi 
e^{- \psi^{+} \Gamma^{-1} \psi -J^{+} \psi - \psi^{+} J} 
\en 
the partition function may be rewritten as 
\be \label{13} 
 Z = \int \prod_{\ii} D\psi_{\ii} D\psi_{\ii}^*  e^{ 
\intau (-\frac{2}{E_J} \sumij \psii^* 
\gamma_{\ii \jj}^{-1} \psij )} e^{-S_{Eff}[\psi]},
\en 
where the effective action for the auxiliary Hubbard-Stratonovich 
field $\psi_\ii$, $S_{Eff}[\psi]$, is given by 
\be
\label{14}
S_{Eff}[\psi]= - \log  \grs \intphi \exp \bigg\{ \intau \big[
-\frac{1}{2}\sumij C_{\ii \jj}\frac{\dot{\varphii}}{2e} 
\frac{\dot{\varphij}}{2e} +i \sumi \big( q_{\ii}
\frac{\dot{\varphii}}{2e}-\psii e^{i\varphii}-\psii^* 
e^{-i \varphii}\big)\big] \bigg\} \grd .
\en 
The Hubbard-Stratonovich field $\psi_{\ii}$ may be regarded as the order 
parameter for the insulator-superconductor phase transition since it
turns out to be proportional to 
$\langle e^{i\varphi_{\ii}}\rangle $, as it can be easily seen from the
classical equations of motion.
From Eq. (\ref{14}), the Ginzburg-Landau free-energy may be derived
by integrating out the phase field $\phi_{\ii}$. 
 
Since the superconductor-insulator phase transition is expected
to be second order \cite{simkin96}, it may be safely assumed that, close
to the onset of superconductivity, the order parameter $\psi_{\ii}$ is
small. One may then expand the effective action up to the second order
in $\psi_{\ii}$, getting
\be \label{15}
S_{Eff}[\psi]= S_{Eff}[0]+\intau \intau' 
G_{\rr \es}(\tau,\tau')\psir(\tau) \psis^*(\tau')+ \cdots \quad ,
\en 
where $G_{\rr \es}$ is the phase correlator 
\be \label{16} 
G_{\rr \es}(\tau, \tau')= \frac{1}{\beta^2}\, \frac{
\delta^2S_{Eff}[\psi]}{\delta \psir(\tau) \delta \psis (\tau')} 
\Bigg|_{\psi,\psi^* = 0} =\langle e^{i\varphir(\tau) 
-i \varphis(\tau')} \rangle_0. 
\en 
The partition function (\ref{13}) can then be written as
\cite{grignani00}
 \be  \label{17}
Z =\int \prod_{\ii} D\psi_{\ii} D\psi_{\ii}^* e^{- \beta F[\psi]} ,
\en 
where $F[\psi]$ is the Ginzburg-Landau free energy; 
due to Eq. (\ref{15}), up to the second order in $\psi_\ii$,
one has
\be
\label{18}
 F[\psi]=\intau \intau'  \sumij 
\psii^*(\tau) [ \gamma_{\ii \jj}^{-1} \delta(\tau-\tau')- 
G_{\ii \jj}(\tau,\tau')]\psij(\tau'). 
\en 
In order to compute the phase correlator $G_{\rr \es}$ one should
evaluate the expectation value in Eq. (\ref{17}) by means of the path
integral over the phase variables $\varphi_{\ii}(\tau)$. In performing
the integration needed for the explicit evaluation of Eq. (\ref{18}),
one should take into account that the field configurations satisfy \be
\label{19} \varphi_{\ii} (\beta)-\varphi_{\ii} (0)=2\pi n_{\ii} .
\en 
It turns out very convenient to untwist the boundary conditions by
decomposing the phase field in terms of a periodic field
$\phi_{\ii}(\tau)$ and a term linear in $\tau$ which takes into account
the boundary conditions (\ref{19}); namely, one sets
\be \label{20} 
\varphi_{\ii}(\tau)=\phi_{\ii}(\tau)+\frac{2\pi}{\beta}n_{\ii}\tau\ , 
\en 
with $\phi_{\ii}(\beta)=\phi_{\ii}(0)$. 
Summing over all the phases $\varphi_{\ii}(\tau)$ amounts then to 
integrate over the periodic field $\phi_{\ii}$ and to sum over the 
integers $n_{\ii}$. 
As a result \cite{grignani00}, the phase correlator factorizes as the
product of a topological term depending on the integers $n_{\ii}$ and a
nontopological one; namely,
$$ 
G_{\bf r s}(\tau;\tau') = 
\frac{\int D \phi_{\ii} e^{i\phi_{\bf r}(\tau)- 
i\phi {\bf_s}(\tau')} \exp \grs \intau (-\frac{1}{2}C_{\bf ij} 
\dot{\frac{\phii}{2e}}\dot{\frac{\phij}{2e}})\grd} 
{\int D \phi_{\ii} \exp \grs \intau (-\frac{1}{2}C_{\bf ij} 
\dot{\frac{\phii}{2e}}\dot{\frac{\phij}{2e}})\grd} \cdot $$ 
\be \label{21} 
\cdot \> \frac{ 
\sum_{[n_{\ii}]} 
e^{i\frac{2 \pi}{\beta}(n_{\rr}\tau-n_{\es}\tau ')} 
e^{\grs \sum_{\bf i j}-\frac{\pi^2}{2 \beta e^2} C_{\bf ij} 
n_{\ii}n_{\jj}+ \sum_{\ii}2i \pi   \frac{q_{\ii}}{2e}n_{\ii} \grd } } 
{  \sum_{[n_{\ii}]} e^{\grs \sum_{\bf i j}-\frac{\pi^2}{2 \beta e^2}C_{\bf ij} 
n_{\ii}n_{\jj}+\sum_{\ii}2i\pi \beta 
\frac{q_{\ii}}{2e}n_{\ii} \grd }}\ . 
\label{corr1} 
\en
For the sake of clarity, we briefly outline the computation
of the factors appearing in the right-hand side
of the previous equation. Firstly, one should compute the path
integral
\be
\label{appcor}
 \frac{\int D \phi_{\ii} e^{i\phi_{\bf
r}(\tau)-i\phi_{\bf s}(\tau')} \exp \grs \intau (-\frac{1}{2}C_{\bf ij}
\frac{\dot{\phii}}{2e}\frac{\dot{\phij}}{2e})\grd} 
{\int D \phi_{\ii} \exp \grs \intau (-\frac{1}{2}C_{\bf ij} 
\frac{\dot{\phii}}{2e}\frac{\dot{\phij}}{2e})\grd}\ . 
\en 
Fourier transforming $\phii(\tau)$ according to 
\be 
\phii(\tau)=\frac{1}{\beta}\sum_{n=-\infty}^{+\infty} 
\phi_{\ii, m}e^{i \omega_{m}\tau} 
\en 
with $0 \le \tau \le \beta$ and $\omega_m=\frac{2\pi}{\beta}m$, 
the numerator of Eq. (\ref{appcor}) becomes
$$ 
\int\prod_{\ii}d\phi_{\ii,0} 
\prod _{n=1}^{\infty}d\phi_{\ii,n}d\phi_{\ii,n}^* 
\exp \bigg\{-\frac{1}{4e^2\beta}\sum_{\ii \jj}\sum_{n=1}^{+\infty} 
C_{\ii \jj}\omega_n^2\phi_{\ii,n} \phi_{\jj,n}^*$$
\be \label{intermed} 
+\frac{i}{\beta}\sum_{n=1}^{\infty}\left(\phi_{\rr,n}e^{i \omega_n \tau}- 
\phi_{\es,n}^*e^{-i \omega_n \tau'}\right)
+ \frac{i}{\beta}\left
(\phi_{\rr,0}-\phi_{\es,0}\right) +c.c. \bigg\}\ .
\en 
Upon integrating over the components $\phi_{\rr,0}$, $\phi_{\es 0}$
one gets a factor $\delta_{\rr \es}$ 
\be
\label{Kappa}
\bigg(\prod_{\ii\ne\rr,\es}\int_{-\infty}^{\infty}d\phi_{\ii,0}\bigg) 
\bigg(\int_{-\infty}^{\infty}d\phi_{\rr,0}\int_{-\infty}^{\infty}d 
\phi_{\es,0}e^{\frac{i}{\beta}
(\phi_{\rr,0}-\phi_{\es,0})}\bigg)= 
\delta_{\rr \es} \cdot K 
\en 
where $K$ is an irrelevant divergent constant which cancels 
against the denominator. 
Using Eq. (\ref{Kappa}), Eq. (\ref{intermed}) becomes
$$ 
K \delta_{\rr \es}\prod_{n=1}^{\infty} 
\int_{-\infty}^{\infty}\prod _{\ii}d\phi_{\ii,n}d\phi_{\ii,n}^* 
\exp\left\{-\frac{1}{4e^2\beta}\sum_{\ii \jj}
C_{\ii \jj}\omega_n^2\phi_{\ii,n} \phi^*_{\jj,n} 
\right.  $$
$$   +\left.\sum_{\ii}\phi_{\ii,n}
\frac{i}{\beta}\delta_{\rr\ii}(e^{i \omega_n \tau}- 
e^{i \omega_n \tau'})-\sum_{\ii}\phi^*_{\ii,n} 
\delta_{\rr \ii}\frac{i}{\beta} 
(e^{-i \omega_n \tau'}-e^{-i \omega_n \tau'})\right\}.
$$ 
The multiple Gaussian integral is easily computed to give \cite{grignani00},
up to an irrelevant constant which cancels against the denominator, 
$$ 
\delta_{\rr \es}\prod_{\ii}\prod_{n=1}^{\infty}\int_{-\infty}^{\infty} 
d\phi_{\ii n} d\phi^*_{\ii n}\exp \bigg\{\sum_{\ii\jj} 
\frac{i}{\beta}\delta_{\rr\ii}(e^{i \omega_n \tau}- 
e^{i \omega_n \tau'}) \cdot$$ 
$$\cdot \big(\frac{4e^2\beta C_{\ii \jj}^{-1}}{\omega_n^2}\big) 
\frac{i}{\beta}\delta_{\rr\ii}(e^{-i \omega_n \tau}- 
e^{-i \omega_n \tau'})\bigg\} 
=$$
$$ 
= \delta_{\rr \es} \exp\bigg\{\frac{8 e^2C_{\rr \rr}^{-1}}{\beta } 
\sum_{n=1}^{\infty}(\frac{1-\cos 
\omega_n(\tau-\tau' )}{\omega_n^2}) \bigg\}
$$ 
where $ -\beta \le \tau-\tau' \le \beta$. By using the identity
\be 
|x|-\frac{x^2}{\beta}=
\sum_{n=1}^{\infty}\left(\frac{4}{\beta \omega_n^2}-
\frac{4\cos \omega_n x}{\beta \omega_n^2}\right) \hspace{1cm}
-\beta \le x \le \beta  ,
\en  
the first term of the right-hand side of Eq. (\ref{corr1}) is found
to be
\be
\label{22}
\delta_{\rr \es} \exp \grs -2e^2C^{-1}_{\rr \rr}\biggl(|\tau-\tau'|- 
\frac{(\tau-\tau')^2}{\beta} \biggr)\grd . 
\en 

The sum over the integers in the topological factor in Eq. (\ref{corr1})
is carried out by means of the Poisson resummation formula
\cite{lebellac}
\be
|\det G|^{\frac{1}{2}} \sum_{[n_{\ii}]} 
e^{ - \pi (n-a)_{\ii}G_{\ii \jj} 
(n-a)_{\jj}}= 
\sum_{[m_{\ii}]} e^{ - \pi m_{\ii}(G^{-1})_{\ii\jj}m_{\jj} 
-2\pi i m_{\ii}a_{\ii}} \label{PR}\ . 
\label{poisson}
\en
Due to Eq. (\ref{poisson}), Eq. (\ref{corr1}) becomes
$$ 
G_{\rr \es}(\tau,\tau')=\delta_{\rr \es}e^{ -2e^2C^{-1}_{\rr \rr} 
|\tau-\tau'|} \>\> \cdot $$ 
\be \label{corr2} 
\cdot \> \frac{\sum_{[n_{\ii}]}e^{  -\sum_{\ii \jj} 2e^2\beta 
C_{\ii \jj}^{-1}(n_{\ii}+\frac{q_{\ii}}{2e})(n_{\jj}+ 
\frac{q_{\jj}}{2e})-\sum_{\ii}4e^2C_{\rr \ii}^{-1}(n_{\ii}+ 
\frac{q_{\ii}}{2e})(\tau-\tau')}  } 
{\sum_{[n_{\ii}]}e^{\sum_{\ii \jj} 2 
\beta e^2C_{\ii \jj}^{-1}(n_{\ii}+\frac{q_{\ii}}{2e})(n_{\jj}+ 
\frac{q_{\jj}}{2e}) }  } 
\en 
with $n_{\ii}$ taking all integer values and $\sum_{[n_{\ii}]}$ being a
sum over all the configurations.

By means of a Euclidean-time Fourier transform, the fields $\psi_{\ii}$
are defined as $$
\psi_{\ii}(\tau)=\frac{1}{\beta}\sum_{\mu} 
\psi_{\ii}(\omega_{\mu})e^{i\omega_{\mu}\tau} \, 
$$ 
where $\omega_\mu$ are the Matsubara frequencies.  As a consequence, the
phase correlator $G_{\ii\jj}$ may be written as
\be \label{24} 
G_{\ii\jj}(\tau;\tau')=\frac{1}{\beta}\sum_{\mu \mu'} 
G_{\ii\jj}(\omega_{\mu} ;\omega_{\mu'})e^{i\omega_{\mu}\tau} 
e^{i\omega_{\mu'}\tau'}\ . 
\en 
From Eq. (\ref{corr2}) one can easily show that
$G_{\rr \es}(\omega_{\mu};\omega_{\mu}')$ 
is diagonal in the Matsubara frequencies and is given by
\be \label{25} 
G_{\rr \es}(\omega_{\mu};\omega_{\mu}')= 
G_{\rr }(\omega_{\mu})\cdot 
\delta_{\rr \es}\cdot\delta(\omega_{\mu}+\omega_{\mu'}) 
\en 
with 
\be
\label{26}
G_{\rr}(\omega_{\mu})=\frac{1}{2E_{c}}\cdot 
\sum_{[n_{\ii}]}  \frac{  \;  e^{-\frac{4}{y} 
\sum_{\ii \jj}\frac{U_{\ii \jj}}{U_{\oo\oo}} 
(n_{\ii}+\frac{q_{\ii}}{2e})(n_{\jj}+\frac{q_{\jj}}{2e})}} 
{ 1-4[\sum_{\jj}\frac{U_{\bf r j}}{U_{\oo \oo}} (n_{\ii}+ 
\frac{q_{\ii}}{2e})-i \omega_{\mu}]^2}\cdot\frac{1}{Z_0} .
\en 
In Eq. (\ref{26}) $Z_0$ is given by
$$ Z_0=  \sum_{[n_{\ii}]}e^{-\frac{4}{y} \sum_{\ii 
\jj}\frac{U_{\ii \jj}}{U_{\oo \oo}} 
(n_{\ii}+\frac{q_{\ii}}{2e})(n_{\jj}+\frac{q_{\jj}}{2e})}
$$ 
with 
$U_{\ii \jj}=4 e^2 C_{\ii \jj}^{-1}$, $ E_C=e^2C_{\rr \rr}^{-1}/2$ and
$y=k_B T_c/E_C$. 
In terms of Matsubara frequencies the Ginzburg-Landau 
free energy (\ref{18}) becomes \cite{panyukov89,vanotterlo93,grignani00}
\be
\label{27}
F[\psi]=\frac{1}{\beta} \sum_{\mu  \ii \jj} 
\psi_{\ii}^*(\omega_{\mu}) 
\biggl[ \frac{2}{E_J}\gamma_{\ii \jj}^{-1}-G_{\ii} 
(\omega_{\mu})\delta_{\ii \jj} \biggr]\psi_{\jj}(\omega_{\mu})\ . 
\en 
Equation (\ref{27}) is the pertinent starting point for the
analysis of the phase boundary between the insulating and the
superconducting phases in JJA with arbitrary capacitance matrix and with
a generic charge frustration.
 
\section{Mean Field Results} 
In the following we shall illustrate the steps involved in the
derivation \cite{grignani00} of the equation determining the phase
boundary in the plane $(\alpha, K_B T_c/E_C)$, in mean-field theory and
for a system with arbitrary capacitance matrix and a uniform
distribution of offset charges. For this purpose, it is convenient to
expand the fields $\psi_{\ii}(\omega_\mu)$ and $G_{\ii}(\omega_\mu)$ in
terms of the vectors of the reciprocal lattice $\kk$. One has
\begin{eqnarray}
\psi_{\ii}(\omega_{\mu})&=&\frac{1}{N}\sum_{\kk}
\psi_{\kk}(\omega_{\mu})e^{i\kk\cdot \ii}\label{28}\\
G_{\ii}(\omega_{\mu})&=&\frac{1}{N}\sum_{\kk}
G_{\kk}(\omega_{\mu})e^{i\kk\cdot \ii}\label{giq}\ . \end{eqnarray}
Moreover 
\be \label{29} 
\gamma_{\ii \jj}^{-1}=\frac{1}{N}\sum_{\kk} 
\gamma_{\kk}^{-1}e^{i\kk\cdot( \ii-\jj)},
\en 
where $\gamma_{\kk}^{-1}$ is the inverse of  the Fourier transform 
of the Josephson coupling strength $\gamma_{\ii\jj}$. As a consequence
$$
\gamma_{\kk}^{-1}=\frac{1}{\sum_{\pp}e^{-i \kk\cdot \pp}} ,
$$ 
where $\pp$ is a vector connecting two nearest neighbors sites. 
Expanding in $\kk$, one gets
\be\label{gammaq} 
\gamma_{\kk}^{-1}=\frac{1}{z}+\frac{{\kk}^2 a^2}{ z^2}+\cdots 
\en 
where $a$ is the lattice spacing.
The first term in Eq. (\ref{gammaq}) provides the mean field theory
approximation which, as expected, is exact in the limit of large 
coordination number. 
 
The Ginzburg-Landau free energy (\ref{27}) reads
\be 
F[\psi]= 
\frac{1}{\beta N}\sum_{\mu \kk\kk'}\psi_{\kk}(\omega_{\mu})^* 
\left[\frac{2}{E_J}\gamma_{\kk}^{-1}\delta_{\kk\kk'}-\frac{G_{\kk-\kk'}
(\omega_{\mu})}{N}\right]\psi_{\kk'}(\omega_{\mu}).
\label{ginzburg-landau-mf} 
\en 
Using Eq. (\ref{gammaq}) and keeping only terms of zero-th order in
$\omega_\mu$ and $\kk$, one obtains the mean field theory approximation
to the coefficient of the quadratic term of $F$:
\be
\label{30}
F[\psi] \simeq \frac{1}{\beta N}\sum_{\kk \mu}
\bigg[ \frac{2}{E_J z}-G_{\bf 0}(0) 
+\cdots \bigg] |\psi_{\kk}(\omega_{\mu})|^2\ . 
\en 
The equation for the phase boundary line then reads as 
\be \label{eqgenlin} 
1=z\frac{E_J}{2}G_{\bf 0}(0) ,
\en 
with 
\be \label{Gconr} 
G_{\bf 0}(0)= 
\frac{1}{N}\sum_{{\bf r}}G_{{\bf r}}(\omega_{\mu}=0,T=T_c). 
\end{equation}

As evidenced in Ref.\cite{grignani00}, the MF theory approximation
amounts to neglect all the higher order terms in Eq.
(\ref{ginzburg-landau-mf}). Equation (\ref{eqgenlin}) determines the 
relation between $T_c$ and $\alpha$ at the phase boundary within the MF
approximation.

For a uniform distribution of offset charges Eq.(\ref{eqgenlin})
simplifies further since in Eq. (\ref{Gconr}) $G_{\rr}$ does not depend
on $\rr$. As a consequence, the phase boundary equation becomes
\be \label{lineacritica} 
1=\alpha\cdot   \sum_{[n_{\ii}]}  \frac{  \;  e^{-\frac{4}{y} \sum_{\ii \jj} 
\frac{U_{\ii \jj}}{U_{\oo\oo}}  (n_{\ii}+\frac{q}{2e})(n_{\jj}+ 
\frac{q}{2e})}} { 1-4[\sum_{\jj}\frac{U_{\bf 0 j}}{U_{\bf 00}} 
(n_{\ii}+\frac{q}{2e})]^2} 
\cdot\frac{1}{Z_0} 
\en 
with 
$$ 
\alpha=\frac{zE_J}{4E_{c}}
$$
and
$$
Z_0=  \sum_{[n_{\ii}]}e^{-\frac{4}{y} \sum_{\ii \jj}
\frac{U_{\ii \jj}}{U_{\bf 00}}(n_{\ii}+\frac{q}{2e})(n_{\jj}+\frac{q}{2e})}. 
$$ 
In the following we shall illustrate through examples the physical
implications of Eq. (\ref{lineacritica}).

\subsection{Self Charging Model} 
For a diagonal capacitance matrix, $U_{\ii \jj}= \delta_{\ii \jj}U_{\oo
\oo}$, one singles out only the self-interaction of plaquettes.
Eq. (\ref{lineacritica}) becomes
\be 
\label{casodiag}
\frac{1}{\alpha}= g(q,y)
\en
where
\begin{equation}
g(q,y)=\frac{\sum_{n}e^{-\frac{4}{y}(n+q/2e)^{2}}\frac{1}{1-4(n+q/2e)^{2
}}} {\sum_{m}e^{-\frac{4}{y}(m+q/2e)^{2}}}. \label{g_qy}
\end{equation}
We observe that Eq. (\ref{casodiag}) is invariant under $q \to q+2e$
and symmetric around $q=n+e$, where $n$ is an integer:
\begin{equation}
g(q+2ne,y)=g(q,y); \quad \quad
g\left[(2ne+e)+q,y\right]=g\left[(2ne+e)-q,y\right].
 \label{per}
\end{equation}

In Fig. 1 we plot $T_c$ as a function of $\alpha$.
One sees that there is no superconductivity for $\alpha<1$.
Due to the periodicity of Eq. (\ref{casodiag}) this holds for any
integer $q$. For $q=e$,  one gets
\be
\label{10}
\alpha=\frac{e^{-\frac{1}{y}}+\sum_{n=1}^{+\infty}
e^{-\frac{4}{y}(n+\frac{1}{2})^2}}{\frac{4+y}{4y}e^{-\frac{1}{y}}
+\sum_{n=1}^{+\infty}\frac{1}{1-4(n+\frac{1}{2})^2}
e^{-\frac{4}{y}(n+\frac{1}{2})^2} }.
\en
From Fig. 2, in which we plot $T_c$ vs. $\alpha$, one sees that
superconductivity is attained for all the values of $\alpha$, since
the superconducting order 
parameter at zero temperature is different from zero:
a uniform offset charge $q=e$ always {\em favors} superconductivity.
In Fig. \ref{fig3} we plot the phase boundary line as a
function of the charge frustration, for a finite value of the critical
temperature $T_c$ and also for $T_c \to 0$.
We observe that the limit $T_c \rightarrow 0$ is singular; nevertheless,
taking this limit in the finite temperature Eq. (\ref{eqgenlin}), one
finds an expression for the phase boundary line in good agreement with
the results reported in Refs. \cite{bruder92,vanotterlo93,kampf93}.

\subsection{Nondiagonal Capacitance Matrices}
In this Section we analyze the situation arising when the diagonal
interaction matrix element $U_{\oo\oo}$ and the nearest neighbor
interaction matrix element $U_{\oo \pp}= \theta \, U_{\oo\oo}$ are
nonzero. To do this, one should expand the critical line Eq.
(\ref{lineacritica}) for $q=0$ and small critical temperatures
\cite{fishman88}: $$
\alpha=1+\big[\frac{8}{3}+2z(1-\frac{1}{1-4\theta^2})\big]
e^{-\frac{4}{y}}+... \quad .
$$
When $\theta>1/\sqrt{4+3z}$, the coefficient of the exponential
$e^{-4/y}$ is negative and the phase boundary line $\alpha=\alpha(T_c)$
first bends to the left;
when the critical temperature is high enough, it bends to the right,
favoring the insulating phase. This is an indication for the
possibility of observing {\em reentrant superconductivity} in these
systems: fixing the parameter of the JJA (i.e., fixing $\alpha$), by
decreasing the temperature it is possible to go down from an insulating
state to a superconducting one and then, further decreasing $T_c$, to go
back to the insulator.

As evidenced by Fishman and Stroud \cite{fishman88}, the regime of
physical interest is $\theta <1/z$; namely, when the capacitance
matrix is invertible. Therefore, in dimensions $D \ge 2$, reentrance
shall not occur with $q=0$ and nearest neighbor interaction matrix
\cite{fishman88}. With $q=e$ and a nearest neighbor interaction matrix,
the situation is different \cite{grignani00}.
We define
\be 
E[n_{\ii}]= \sum_{\ii\; \jj}
\frac{U_{\ii \jj}}{U_{\bf 00}}({n}_{\ii}+\frac{1}{2})(n_{\jj}+\frac{1}{2})  
\en
the energy of a generic charge distribution on the lattice $\{ n_{\ii} \}$.
Denoting with $n^0_{\ii}$ and $n^1_{\ii}$ the charge distributions of
the two lowest lying energy states and with $E^0$ and $E^1$ the
corresponding energies, the low temperature expansion of
Eq. (\ref{lineacritica}) yields \be
\label{expans}
\alpha=
\frac{ \sum_{[n^0]}e^{-\frac{4}{y}E^0}+\sum_{[n^1]}e^{-\frac{4}{y}E^1}
+\cdots}{   \sum_{[n^0]}
\frac{e^{-\frac{4}{y}E^0}}{1-4\big[\sum_{\jj}
\frac{U_{\oo\jj}}{U_{\oo\oo}}(n_{\jj}^0+\frac{1}{2})\big]^2}
+\sum_{[n^1]}\frac{e^{-\frac{4}{y}E^1}}{1-4\big[\sum_{\jj}
\frac{U_{\oo\jj}}{U_{\oo\oo}}(n_{\jj}^1+\frac{1}{2})\big]^2}+\cdots }\ .
\en

Independently on the explicit form of $U_{\ii\jj}$,
$E[n_{\ii}]$ (for a square lattice in $D$ dimensions) reaches its
minimum value when
$(n_{\ii}^0+\frac{1}{2})=\pm\frac{1}{2}(-1)^{i_1+i_2+...+i_D}$ with
$i_j\ (j=1,...,D)$ the components of the lattice position vector $\ii$
in units of the lattice spacing.

For models with nearest-neighbor interaction, i.e., 
$U_{\ii\jj}/U_{\oo \oo}=\delta_{\ii\jj} +
\theta\sum_{\pp}\delta_{\ii+\pp,\jj}$
with $\sum_{\pp}$ denoting summation over nearest neighbors,
the first excited state 
has an energy 
$E[n_{\ii}^1]=E[n_{\ii}^0]+z\theta$, where $E[n^0_{\ii}]$, 
the ground state energy, is given by
$\sum_{\ii}(1-z\theta)/4$.

\begin{figure}[h]
\centerline{\psfig{figure=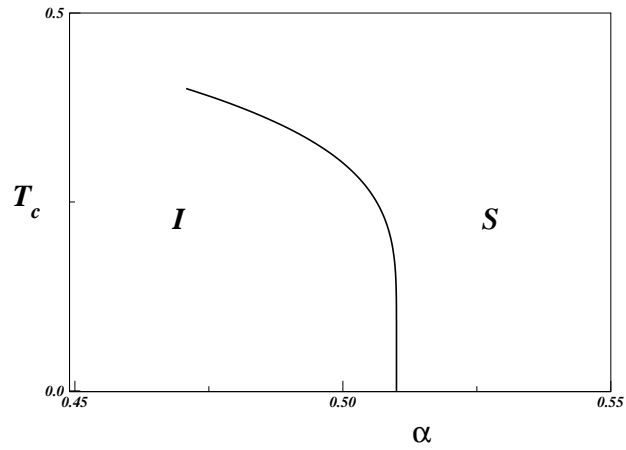,width=72mm,angle=270}}
\caption{Low temperature expansion of the critical line with a
short-ranged inverse capacitance matrix. In the plot, $z=6$ and
$\theta=0.05$, where $\theta$ is the ratio between nearest-neighbor and
diagonal terms.}
\label{fig4} \end{figure}

With the above values of $E[n_{\ii}^0]$ and $E[n_{\ii}^1]$
and keeping only the leading order term in $T_c$,
Eq. (\ref{expans}) becomes
\be \label{moti2} 
\alpha=(1-(1-z\theta)^2) \cdot
(1+ a_1 e^{-\frac{4}{y}z\theta}+
\cdots )
\en
where
\be \label{moti}
a_1 \equiv \bigg(1-\frac{1-(1-z\theta)^2}{1-(1+z\theta)^2}\bigg)
+z\bigg(1-\frac{1-(1-z\theta)^2}{1-(1-(z-2)\theta)^2}\bigg).
\en
Reentrant behavior at low temperature occurs
when the coefficient 
of the exponential is negative, namely when $a_1<0$. In
Fig. \ref{fig4} we plot $T_c$ versus $\alpha$ for $\theta=0.05$ and
$z=6$. The resulting diagram exhibits reentrance in the
insulating phase 
even for models with nearest neighbors interaction.  For a detailed
study of the lobe diagrams in JJA with non-diagonal capacitance
matrices, see \cite{fazio01}.

\section{Capacitive disorder}

In this Section, we shall determine the finite
temperature phase diagram of JJA with capacitive disorder (i.e., with random 
offset charges and/or random self-capacitances). To derive the phase 
boundary between the insulating and the superconducting phase, we shall use 
the path-integral approach for quantum JJA
with offset charges and general capacitance matrices reviewed 
in the previous Sections. We find that charge disorder supports
superconductivity and that the relative variations of 
the insulating and superconducting regions 
depend on the mean value $q$ of the charge probability distribution: when 
$q=0$, increasing the disorder leads to an enlargement of 
the superconducting phase. If the charge disorder is sufficiently
strong ($\sigma \gtrsim e$), the lobe structure \cite{fazio01}
disappears: in other words, the phase boundary line (and the correlation
functions) do not depend any longer on $q$.
In the following, we shall provide a quantitative analysis of this 
phenomenon. Also the randomness of
the self-capacitances leads to remarkable effects, namely, the
superconducting phase increases with respect to the case where disorder
is not present.

We shall consider several probability distributions which we expect to
provide a realistic description of experimental situations.
The low temperature behavior obtained by a pertinent extrapolation of
our finite $T$ results is consistent with the phase diagram obtained in
Ref. \cite{fisher89}.

For a given realization of the disorder, the Ginzburg-Landau free energy
(i.e., the free energy near the transition) is given by Eq. (\ref{30}).
We shall perform a quenched average, in which each of the random
variables takes a unique value as the statistical variables fluctuate.
This corresponds to taking the average of the logarithm of the partition
function, i.e., the free energy. The average of  the free energy over
all the possible realization of the disorder allows for the evaluation
of the effect of a random charge frustration $\{q_{{\bf i}}\}$ or a
random diagonal charging energy terms $U_{{\bf i} {\bf i}}=4e^2 C_{{\bf
i}{\bf i} }^{-1}$. The pertinent starting point for the analysis of
these situations is then
\begin{equation}
 \label{av-ginzburg-landau}
\bar{F}[\psi]=\int d\{X\} P(\{X\}) F[\psi] 
\end{equation} 
where $P(\{X\})$ is a given probability distribution and $d\{X\} 
P(\{X\})=\prod_{{\bf i}} dq_{{\bf i}} P(q_{{\bf i}})$ if one
analyzes the effect of random offset charges or $d\{X\}
P(\{X\})=\prod_{{\bf i}} dU_{{\bf i}{\bf i} } P(U_{{\bf i}{\bf i}})$ for
random charging energies. The random variables on different sites are
taken to be independent. The phase boundary line between the insulating
and the superconducting phase is determined by requiring that
$\bar{F}=0$, which in turn leads to \cite{mancini03}
\begin{equation}
\label{boundary-line}
1=z\frac{E_J}{2}
 \bar{G_{{\bf 0}}}.
\end{equation} 
 
\subsection{Random offset charges} 
 
In the following, we shall consider three different random offset charges 
probability distribution with mean $q$ and width $\sigma$. That is, a
Gaussian distribution $P(q_{{\bf i}})= const \cdot e^{-(q_{{\bf i}
}-q)^2/2\sigma^2}$, a uniform distribution $P(q_{{\bf i}})= const$
between $ q-\sigma$ and $q+\sigma$ and $0$ otherwise,
and a sum of $\delta$-like distributions $P(q_{{\bf i}})=\sum_n p_n
\delta(q_{{\bf i}}-ne)$, with $\sum_n p_n=1$. For a diagonal capacitance
matrix, Eq. (\ref{boundary-line}) leads to
\begin{equation}
\frac{1}{\alpha }=\int
dq P(q)g(q,y)
\label{boundary-line-charge}
\end{equation}
with $\alpha =zE_{J}/4E_{c}$, $y=k_{B}T_{c}/E_{c}$ and $g(q,y)$ given by
Eq. (\ref{g_qy}). If one considers the infinite-range hopping limit, one
still gets Eq. (\ref {boundary-line-charge}) \cite{mancini03}.

\begin{figure}[h] 
\centerline{\psfig{figure=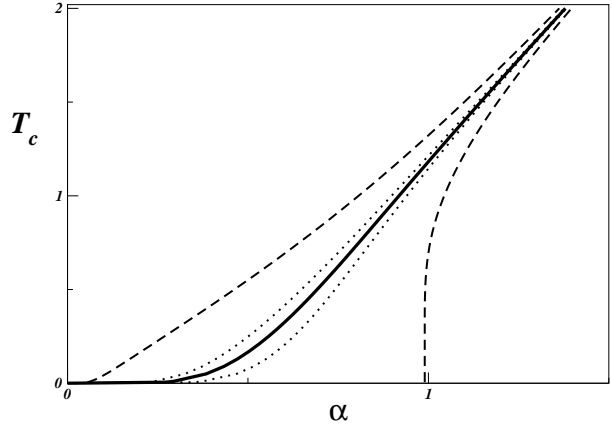,width=72mm,angle=270}}
\caption{Critical line for random offset charges with Gaussian
distribution ($T_c$ is in units of $k_B /E_C$). The bold line is for 
$\sigma=e$: for $\sigma \protect\gtrsim e$ no significant deviations
from this line are observed for all the values of $q$. To the left
(right), we plot $q=e$ ($q=0$); in the plot $\sigma/2e=0.05$ (dashed
lines) and $0.3$ (dotted lines).}
 \label{fig5}
\end{figure}

The results obtained from Eq. (\ref{boundary-line-charge}) with a
Gaussian distribution are displayed in Fig. \ref{fig5}. One observes
that, when $q=0$, increasing $\sigma $ favors the superconducting
phase while, when $q=e$, increasing $\sigma $ leads to the
increase of the insulating phase. For large $\sigma$ (i.e. $\sigma
\gtrsim e$), the phase boundary line is the same for all the values of
$q$ (in Fig. \ref{fig5} the large $\sigma$ behavior is represented by
the bold line). This is expected since, when $\sigma$ is large, the
average free energy $\bar{F}$ does not depend any longer on $q$.
 
As seen in the previous Sections an useful representation of the phase
diagram is provided if one plots, at fixed critical temperature, the
phase boundary line on the plane $q-\alpha$. Without disorder one
observes the lobe structure as discussed in Section III. In the presence
of weak disorder and for $T_c \rightarrow 0$, the lobes
shrink, evidencing a decrease of the insulating phase: for a Gaussian
(or unbounded) distribution the insulating phase completely disappears
even for an arbitrarily weak disorder \cite{fisher89}.

\begin{figure}[h] 
\centerline{\psfig{figure=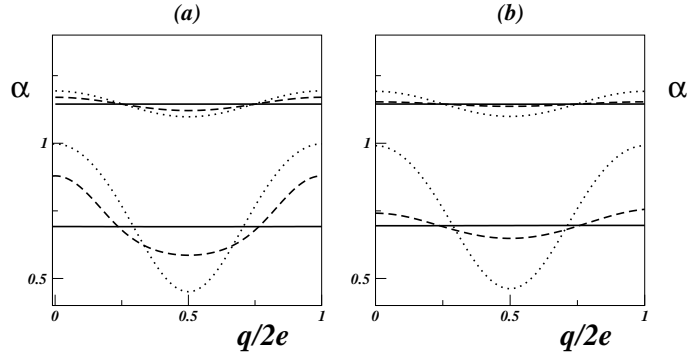,width=72mm,angle=270}}
\caption{Phase diagram with diagonal capacitances and
random offset charges with uniform (a) and Gaussian (b)
distribution. Top (bottom) of the figures: $k_B T /E_C=1.5 (0.5)$. We
plot the cases $\sigma/2e=0.05$ (dotted lines), $0.3$ (dashed
lines)and $0.5$ (solid lines). For large $\sigma/2e$ the phase boundary
line is flat and it is the same for both distributions.}
\label{fig6}
\end{figure}

\begin{figure}[h] 
\centerline{\psfig{figure=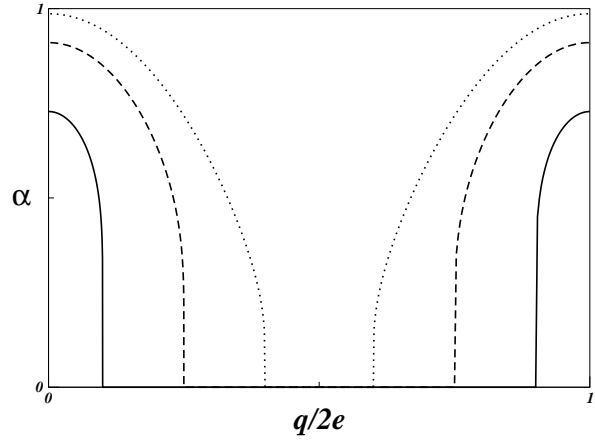,width=72mm,angle=270}}
\caption{Phase diagram at very low critical temperatures for a diagonal
inverse capacitance matrix and random offset charges with uniform
distribution. We plot $\sigma/2e=0.1$ (dotted line), $0.25$ (dashed
line) and $0.40$ (solid line).}
\label{fig7} \end{figure}

In Fig. \ref{fig6} we plot the phase boundary line on the plane
$q-\alpha$ at finite $T_c$ for the Gaussian and uniform distributions
\cite{mancini03}. When the disorder increases the lobes flatten even
at finite temperature and the same lobe structure is obtained from both
distributions. In the limit $T_c \to 0$, one recovers the
result of Ref. \cite{fisher89}. This can be easily seen if one observes
that, at very low temperatures, for $|q| < e$, one has from Eq.
(\ref{g_qy}) \begin{equation} g(q,y \to 0) =
\frac{1}{1-4(\frac{q}{2e})^2}.
\end{equation}
Without disorder ($\sigma=0$), Eq. (\ref{boundary-line-charge}) simply
gives $\alpha=1-4(q/2e)^2$. With the Gaussian distribution, since $g$
has a pole in the half-integer value of the Cooper charge, the integral
in Eq. (\ref{boundary-line-charge}) diverges and $\alpha \to 0$, i.e.,
the lobes disappear for every value of $\sigma$. As evidenced in Fig.
\ref{fig7}, for a uniform distribution, when $\sigma > e$, then $\alpha
\to 0$; when $\sigma < e$, $\alpha\to 0$ only for $e-\sigma \le q \le
e+\sigma$ in agreement with \cite{fisher89}.

\begin{figure}[h] 
\centerline{\psfig{figure=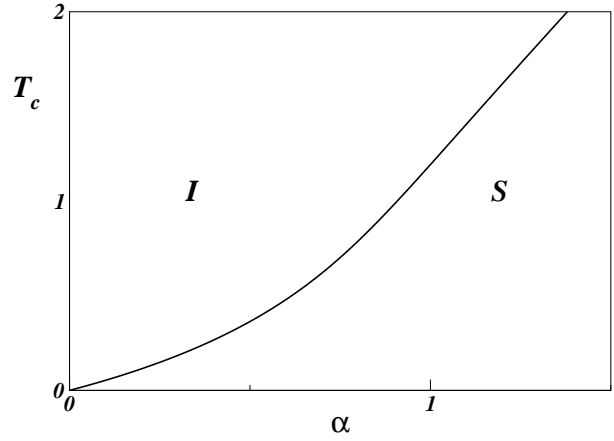,width=72mm,angle=270}}
\caption{Phase diagram for random offset charges with the
probability distribution given by Eq. (\protect\ref{prob_delta_def}) and
diagonal capacitance matrix. In the plot we take $p_0=p_e=1/2$.}
\label{fig8} \end{figure}

Another interesting situation arises if one considers
\begin{equation}
P(q_{{\bf i}})=\sum_n p_n \delta(q_{{\bf i}}-ne),   
\label{prob_delta_def} 
\end{equation} 
with $\sum_n p_n=1$. This corresponds to a random distribution of
charges which are integer multiples of $e$ and, actually, this is the
most realistic situation for a random distribution. In fact, the
probability distributions employed before should be viewed as fictitious
continuous distributions, i.e., the properties of the overall
distribution of charges (mean value and width) can be well approximated
with a continuous distribution $P(q)$. Inserting the probability
distribution (\ref{prob_delta_def}) in Eq. (\ref{boundary-line-charge})
one has
\[
\frac{1}{\alpha}= \int dq
\sum_n p_n \delta(q-ne) g(q,y)=\sum_{odd} p_n g(ne,y)+ \sum_{even} p_n
g(ne,y) ,
 \]
where $\sum_{odd}$ ($\sum_{even}$) is a sum restricted to odd (even) 
integer. From Eq. (\ref{per}), one has $g(2ne,y)= g(0,y)$ and $
g((2n+1)e,y)=g(e,y)$, which leads to
\begin{equation} 
\frac{1}{\alpha}=p_0 \, g(0,y)+p_e\,  g(e,y)  ,
\label{prob_delta}
\end{equation} 
where $p_0=\sum_{even} p_n$ ($p_e=\sum_{odd} p_n$) is the probability that 
the offset charge $q$ is an even (odd) integer multiple of $e$. In Fig.
\ref{fig8} we plot the phase boundary line (\ref{prob_delta}) for
$p_0=p_e=1/2$.

\begin{figure}[h] 
\centerline{\psfig{figure=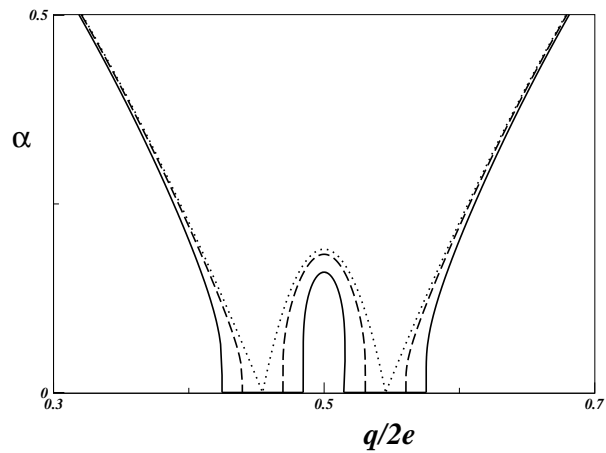,width=72mm,angle=270}}
\caption{Disappearance of the insulating lobe around $q=e$ for a
short-ranged inverse capacitance matrix. The phase diagram is plotted
for $T_c \rightarrow 0$ and with random offset charges uniformly
distributed. In the plot $z\theta$ is equal to $0.1$ while $\sigma/2e$
is respectively $0$ (dotted line), $0.015$ (dashed line) and $0.03$
(solid line). For this value of $z\theta$, the lobe disappears at
$\sigma/2e=0.045$.}
\label{fig9}
\end{figure}

We discuss now the effects induced by nondiagonal capacitance
matrices in the limit $T_c \to 0$ \cite{mancini03}. The phase diagram
without disorder becomes richer \cite{fazio01}; for concreteness, we
shall consider on-site and a weaker nearest-neighbor (NN) interaction,
i.e., the inverse capacitance matrix is restricted to diagonal and NN
terms. If one defines $\theta$ as the ratio between NN and diagonal
terms, one should restrict only to $ z\theta<1$ in order to insure the
invertibility of the capacitance matrix  \cite{fishman88}. Without
disorder, at very low temperatures an insulating lobe around $q=e$
appears \cite{vanotterlo93}: the width of this lobe is
$z\theta/(1+z\theta)$. Putting $ W=1+z\theta$, Eq. (\ref{g_qy}) for
$|q/2e|<1/2W$ gives $g(q,y \to 0)=1/[1-4W^2(q/2e)^2]$; for
$1/2W<q/2e<1-(1/2W)$ it becomes, \begin{equation}
g(q,y \to 0) = -\frac{1}{2}\left[\frac{1}{(2W\frac{q}{2e}-1) (2W\frac{q}{2e}
-3)}+ \frac{1}{(2W(\frac{q}{2e}-1)+1)(2W(\frac{q}{2e}-1)+3)}\right]. 
\end{equation} 
In presence of disorder, Eq. (\ref{boundary-line-charge}) for a uniform
distribution gives $\alpha=0$ for $(1/2W)- \sigma \le q \le (1/2W)+ \sigma$ 
and $1- (1/2W)- \sigma\le q \le 1- (1/2W) + \sigma$. Thus, the lobe width 
decreases as $(z\theta-2\sigma W)/W$. One sees that for $\sigma = z\theta/2W$ 
the insulating lobe around $q=e$ disappears. This phenomenon is evidenced in 
Fig. \ref{fig9}.

\subsection{Random Self-Capacitances}

In many instances, it may happen that the network's parameters are
not uniform across the whole array: despite recent advances in
fabrication techniques, with the use of submicron lithography, variation
of junction parameters associated to the shape of the
islands can be also of $20\%$ \cite{fazio01}. Thus, it is relevant in
many practical situations to study JJA with randomly distributed
self-capacitances: this corresponds to have a random diagonal charging
energy \cite{mancini03,alsaidi03}. In Ref. \cite{alsaidi03} the effects
of disorder on the spectrum of elementary excitations at low
temperatures have been studied.

In this Section we shall study JJA at finite temperature with uniform
charge frustration $q$ and random self-capacitance $C_{{\bf i} {\bf
i}}$, ignoring non-diagonal contributions to the capacitance matrix. The
diagonal charging energy terms $U_{{\bf i} {\bf i}}$ are related to the
self-capacitances $C_{{\bf i} {\bf i}}$ via $U_{{\bf i} {\bf i}}=4e^2
C_{{\bf i} {\bf i}}^{-1}$ and are assumed to be independently
distributed according to the probability distribution $P(U_{{\bf i} {\bf
i}}) \propto e^{-(U_{{\bf i} {\bf i} }-U_0)^2/2\sigma ^2}$. The average
charging energy is defined as $E_C^0= U_0/8$. The diagonal
electrostatic contribution to the energy $U_{{\bf i} {\bf i} }$ needs to
be positive.

\begin{figure}
\centerline{\psfig{figure=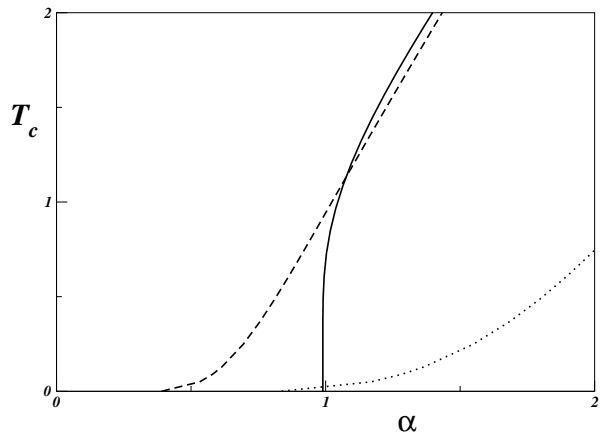,width=72mm,angle=270}}
\caption{Phase diagram in the $T_c-\alpha$ plane for random diagonal 
capacitance with Gaussian distribution and without charge
frustration ($T_c$ is in units of $k_B/U_0$) while $\sigma/2e$ is $0.1$
(solid line), $1$ (dashed line) and $5$ (dotted line). For each line,
the superconducting (insulating) phase lies on the right (left).}
\label{fig10}
\end{figure}

\begin{figure}
\centerline{\psfig{figure=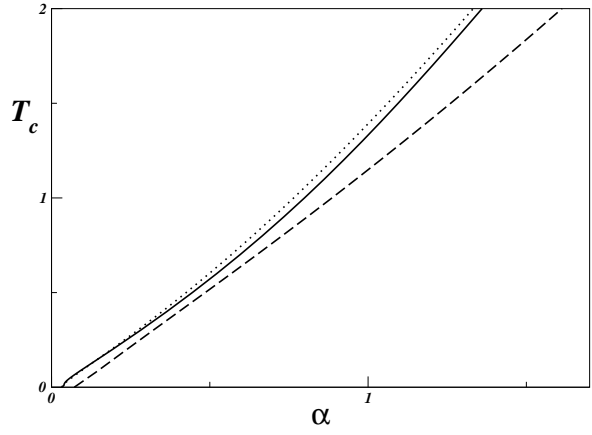,width=72mm,angle=270}}
\caption{Phase diagram in the $T_c-\alpha$ plane for random diagonal
capacitance with Gaussian distribution and uniform offset charge
$q=e$ ($T_c$ is in units of $k_B/U_0$) while $\sigma/2e$ is $0.1$ (solid
line), $1$ (dashed line) and $5$ (dotted line).}
\label{fig11}
\end{figure}

By averaging the free energy (\ref{av-ginzburg-landau}), the equation for
the phase boundary becomes \cite{mancini03}
\begin{equation}
 \label{boundary-line-energy}
\frac{1}{\alpha}=\int_0^{\infty} dU \frac{P(U)}{U} g(U,y) 
\end{equation}
where now  $\alpha=zE_J /4 E_C^0$, $y=k_B T_c / E_C^0$, and
$U=U_{ii}/U_0$; the function $g(U,y)$ is given by
\begin{equation}  \label{g_qU} 
g(U,y)=\frac{\sum_n e^{-\frac{4}{y}U(n+q/2e)^2} \frac{1}{1-4(n+q/2e)^2} }
{\sum_{m}e^{-\frac{4}{y}U(m+q/2e)^2} }.
\end{equation} 
The results of Eq. (\ref{boundary-line-energy}) are summarized in Figs.
\ref{fig10} and \ref{fig11}: when $\sigma $ is small, the
superconducting phase increases in comparison to the situation
in which all self-capacitances are equal: this is due to
the factor $ 1/U$ in Eq. (\ref{boundary-line-energy}), which makes
larger the contribution of junctions with charging energies less than
$U_{0}$. An interesting observation is that, when $q=e$ (maximum
frustration induced by the external offset charges), the randomness does
not modify considerably the phase diagram. This should be compared with
the nonfrustrated case ($q=0$), where randomness sensibly affects the
phase diagram.

\begin{figure}
\centerline{\psfig{figure=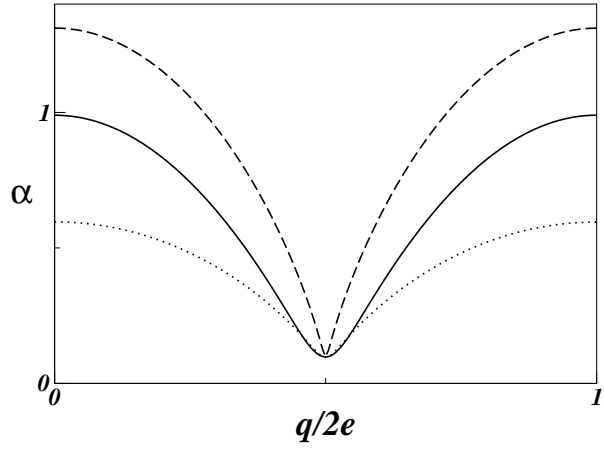,width=72mm,angle=270}}
\caption{Phase diagram in the $q-\alpha$  plane for random diagonal
capacitance with Gaussian distribution and uniform offset charge
at $k_B T /U_0=0.1$. We plot $\sigma/2e=0.1$ (solid line), $1$ (dotted
line) and $5$ (dashed line).}
\label{fig12}
\end{figure}

The increase of the superconducting phase is due to a decrease of
the effective value of the charging energy. This behavior occurs until
$\sigma $ reaches a critical value (depending on the charge frustration
and on the temperature), of order $U_0$: at this value of $\sigma $ the
insulating region starts to increase. This is due to the asymmetry of
the distribution, which has its peak in $U_{0}$, but only for positive
values. This phenomenon is present also if one considers different
distributions and is clearly seen in Fig. \ref{fig12}, where
the phase diagram in the $q$-$\alpha$ plane for different values of the
variance $\sigma$ is plotted.

\section{Concluding Remarks}

In this paper, we reviewed the use of the path integral approach to
finite temperature mean-field theory to analyze the effects induced by
offset charges on the finite temperature phase diagram of Josephson
junction arrays. We provided, for a general Coulomb interaction matrix,
the explicit derivation of the equation for the phase boundary line
between the insulating and superconducting phase.

The resulting phase diagram is drawn in the diagonal case for a generic
uniform offset charge distribution $q$: with $q=e$, the superconducting
phase increases with respect to $q=0$, and the model exhibits
superconductivity for all the values of $\alpha=z E_J/4E_C$. An offset
charge $q=e$ tends to decrease the charging energy and thus favors the
superconducting behavior even for small Josephson energies.

For the model with nearest neighbor inverse capacitance
matrix and uniform offset charge $q=e$, we determined,
in the low temperature expansion, the most relevant contributions
to the equation for the phase boundary. For this purpose we explicitly
constructed the charge distributions on the lattice corresponding to the
lowest energies: a reentrant behavior is found even with a short ranged
interaction.

We also obtained the phase diagram at finite temperature
of JJA with capacitive disorder. For a random distribution of offset
charges with mean $q$ and variance $\sigma$, one has that for $\sigma
\gtrsim e$, the phase boundary line coincides for any value of $q$ and
the lobe structure on the plane $ q-\alpha$ disappears ($\alpha$ is the
ratio between the Josephson and charging energies). For very low
temperatures there is agreement with the result of Ref. \cite{fisher89}.

If one considers also a nearest-neighbor interaction, the
insulating lobe around $q=e$, which arises in absence of disorder, is
destroyed even for small values of $\sigma$. For arrays with random
charging energies, when the variance of the probability distribution is
smaller than a critical value, the superconducting phase increases with
respect to the situation in which all self-capacitances are equal.

It is comforting to observe that the finite temperature MF theory
approach developed in this paper provides results which are in good
agreement with those obtained by recent quantum Monte Carlo simulations
\cite{al-saidi03_2} and by use of improved variational methods
\cite{cooper03}.

{\bf Acknowledgements} We thank  G. Grignani and A. Mattoni for
very stimulating discussions and for their invaluable contributions to
the initial stages of our research. We greatly benefited from
enlightening discussions with S. R. Shenoy and A. Tagliacozzo. We
acknowledge financial support by M.I.U.R. through grant No. 2001028294.

\end{document}